\documentclass[sigconf]{acmart}

\ifdefined\podcbrief
  \copyrightyear{2020}
  \acmYear{2020}
  \setcopyright{acmlicensed}\acmConference[PODC '20]{ACM Symposium on Principles of Distributed Computing}{August 3--7, 2020}{Virtual Event, Italy}
  \acmBooktitle{ACM Symposium on Principles of Distributed Computing (PODC '20), August 3--7, 2020, Virtual Event, Italy}
  \acmPrice{15.00}
  \acmDOI{10.1145/3382734.3405749}
  \acmISBN{978-1-4503-7582-5/20/08}
\else
  \renewcommand\footnotetextcopyrightpermission[1]{}
\fi

\usepackage{subcaption}
\usepackage{tikz}
\usetikzlibrary{arrows,shapes,automata,backgrounds,petri,patterns,positioning}

\usepackage[utf8]{inputenc}
\usepackage{amsfonts}

\usepackage{amsmath}
\usepackage{amsthm}
\usepackage{cleveref}
\newcommand{\n}[1]{\textsf{#1}}
\newtheorem{axiom}{Axiom}
\newtheorem*{theorem*}{Theorem}
\newtheorem{fact}{Fact}
\newtheorem{prop}{Proposition}

\crefname{prop}{proposition}{propositions}

\DeclareUnicodeCharacter{2192}{\ensuremath{\rightarrow}} % →
\DeclareUnicodeCharacter{2208}{\ensuremath{\in}} % ∈
\DeclareUnicodeCharacter{2261}{\ensuremath{\equiv}} % ≡
\DeclareUnicodeCharacter{2264}{\ensuremath{\leq}} % ≤
\DeclareUnicodeCharacter{22C5}{\ensuremath{\cdot}} % ⋅
\DeclareUnicodeCharacter{2080}{\ensuremath{_0}} % ₀
\DeclareUnicodeCharacter{2081}{\ensuremath{_1}} % ₁
\DeclareUnicodeCharacter{2082}{\ensuremath{_2}} % ₂
\DeclareUnicodeCharacter{2124}{\ensuremath{\mathbb{Z}}} % ℤ
\DeclareUnicodeCharacter{3B5}{\ensuremath{\epsilon}} % ε
\DeclareUnicodeCharacter{3C6}{\ensuremath{\phi}} % φ
\DeclareUnicodeCharacter{3C8}{\ensuremath{\psi}} % ψ

\begin{document}
\ifdefined\podcbrief
  \title{Brief Announcement: The Only Undoable CRDTs are Counters}
\else
  \title{The Only Undoable CRDTs are Counters}
\fi
\author{Stephen Dolan}
\email{stedolan@stedolan.net}
\affiliation{OCaml Labs}

\begin{abstract}
In comparing well-known CRDTs representing sets that can grow and
shrink, we find caveats. In one, the removal of an element cannot be
reliably undone. In another, undesirable states are attainable, such
as when an element is present -1 times (and so must be added for the
set to become empty). The first lacks a general-purpose undo, while
the second acts less like a set and more like a tuple of counters, one
per possible element.

Using some group theory, we show that this trade-off
is unavoidable: \emph{every undoable CRDT is a tuple of
  counters}.
\end{abstract}

\maketitle

%% 
%% Notes:
%% 

%% In the infinite case, a reasonable generalisation of the
%% classification of finitely generated abelian groups is this:
%% 
%%   - all abelian groups embed in a divisible group
%% 
%%   - all divisible groups are isomorphic to a direct sum
%%     of Z(p^n) (Prufer groups) and Q

%% 
%% LogRoot-Undo is a reasonable structure - counters for every possible edit.
%% Usually counters are 0 or 1, but Z must be supported for CRDT props.
%% 
%% They note that all operations commute - here's why.
%% 
%% They have a bug in their paper: they don't think that >1 is possible.
%% 

\section{Introduction}

Conflict-free replicated data types (CRDTs) allow replication of a
data structure across multiple machines without risking conflicts
between different versions. Even though each machine may concurrently
modify its own copy of the data structure, a CRDT guarantees that
these concurrent modifications can be merged into a consistent result,
upon which the whole network will agree.

Here, we adopt the \emph{operation-based} view of CRDTs~\cite{shapiro-crdt}, in which a
CRDT consists of some \emph{state} and some \emph{operations}
affecting it, where any two operations that may be performed
concurrently must commute.  If two machines' local replicas go out of
sync by applying different operations concurrently, they can later
merge by exchanging logs of applied operations and applying the other
machine's operations to their own state. The commutativity condition
ensures that both end up in the same final state, despite applying the
operations in different orders.

Below, we review several examples of CRDTs for counters and
sets. For more details, see Shapiro et al.'s comprehensive
survey~\cite{shapiro-comp}.

\subsection{The counter CRDT}

The counter is a simple CRDT, whose state is an integer and whose
operations are increment and decrement. These commute, since $(n + 1)
- 1 = n = (n - 1) + 1$. This ensures that once all machines have seen
all operations, all will agree on the counter's final value.

The counter CRDT is \emph{undoable}: After incrementing we may
decrement to restore the previous state, and likewise we can undo
decrementing by incrementing.

The \emph{modulo-$n$ counter} is a slight variant, where increment
wraps around from $n-1$ to $0$.  Like the ordinary counter,
all operations on the modulo-$n$ counter can be undone.

\subsection{The G-Set CRDT}

Another simple CRDT is the \emph{grow-only set} or G-Set, whose state
is a set of elements and whose operations are $\n{add}\;{\tt A}$ for
each possible element {\tt A}.  Eventual convergence is guaranteed
because $\n{add}\;{\tt A}$ and $\n{add}\;{\tt B}$ commute.

%For any two elements {\tt A},
%{\tt B}, the operations $\n{add}\;{\tt A}$ and $\n{add}\;{\tt B}$
%commute, so again eventual convergence is guaranteed.

However, as the name implies, a G-Set can never shrink.  There is no
$\n{remove}$ operation with which to undo an $\n{add}$, and adding one
turns out to be tricky.  Below, we review several approaches.

% There are several well-known CRDTs
% representing sets that can both grow and shrink, which have subtle
% differences and caveats~\cite{shapiro-comp}. Below we consider
% several.
% 

\subsection{Sets with removal: the OR-Set}

\begin{figure}

\begin{subfigure}[b]{0.45\textwidth}
\emph{Communication between replicas:}

\begin{tikzpicture}[node distance=1.8cm]
\node (s0) {$\{\}$};
\node (s1) [right of=s0] {$\{{\tt A}\}$}
  edge [pre,auto=right] node {$\n{add}_i\;{\tt A}$} (s0);
\node (s2) [above right of=s1] {$\{\}$}
  edge [pre,auto=left] node {$\n{remove}_i\;{\tt A}$} (s1);
\node (s3) [below right of=s1] {$\{\}$}
  edge [pre,auto=right] node {$\n{remove}_i\;{\tt A}$} (s1);
\node (s4) [right of=s2] {$\{{\tt A}\}$}
  edge [pre,auto=right] node {$\textsf{add}_j\;{\tt A}$} (s2);
\node (s') [below right of=s4] {$s$}
  edge [pre,dashed] node {} (s4)
  edge [pre,bend left,dashed] node {} (s3);
\end{tikzpicture}

\emph{Sequences of operations performed to yield state $s$:}
\begin{align*}
S₁ &= \n{add}_i\;{\tt A}; \n{remove}_i\;{\tt A}; \textsf{add}_j\;{\tt A}; \n{remove}_i\;{\tt A} \\
S₂ &= \n{add}_i\;{\tt A}; \n{remove}_i\;{\tt A}; \n{remove}_i\;{\tt A}; \textsf{add}_j\;{\tt A} 
\end{align*}
\caption{First replica removes and re-adds}
\label{remove-undo-example}
\end{subfigure}
\qquad
\begin{subfigure}[b]{0.45\textwidth}
\emph{Communication between replicas:}

\begin{tikzpicture}[node distance=1.8cm]
\node (s0) {$\{\}$};
\node (s1) [right of=s0] {$\{{\tt A}\}$}
  edge [pre,auto=right] node {$\n{add}_i\;{\tt A}$} (s0);
\node (s3) [below right of=s1] {$\{\}$}
  edge [pre,auto=right] node {$\n{remove}_i\;{\tt A}$} (s1);
\node (s') [below right of=s4] {$s'$}
  edge [pre,bend right,dashed] node {} (s1)
  edge [pre,bend left,dashed] node {} (s3);
\end{tikzpicture}

\emph{Sequences of operations performed to yield state $s'$:}
\begin{align*}
S₁ &= \n{add}_i\;{\tt A}; \n{remove}_i\;{\tt A} \\
S₂ &= \n{add}_i\;{\tt A}; \n{remove}_i\;{\tt A}
\end{align*}
\caption{First replica does nothing}
\label{remove-noop-example}
\end{subfigure}

{\footnotesize
Subscripts denote the IDs of $\n{add}$ operations (relevant only for OR-Sets)
}

\caption{Example of undoing $\n{remove}$}
\end{figure}

In an OR-Set (or \emph{add-wins} set~\cite{addwins}) an
element is present if it has been added since it was last removed. We
represent this with two G-Sets, \emph{added} and \emph{removed}, each
containing pairs of an element and an ID. An element ${\tt A}$ is
deemed present in the set if there is some $i$ such that $({\tt A},
i)$ is in the added but not the removed set.

The $\n{add}\;{\tt A}$ operation inserts $({\tt A},
i)$ into the added set (with some fresh ID $i$), and the
$\n{remove}\;{\tt A}$ operation inserts $({\tt A}, j)$
into the removed set, for each $j$ where $({\tt A}, j)$ is in the added set.

This means that each $\n{remove}$ operation undoes all prior
$\n{add}$s. However, undoing a remove is less
straightforward. Consider the example in \cref{remove-undo-example}:
we start with the empty set and add {\tt A} to it, at which point two
replicas diverge. The first removes and then re-adds {\tt A}, while
the second just removes it. Afterwards, the two replicas merge,
yielding state $s$.

In \cref{remove-undo-example}, $S_1$ and $S_2$ describe the sequences
of operations performed by the two replicas after merging. The operations commute,
so both
yield the same final state $s$. In state $s$, both the added and
removed sets contain $({\tt A}, i)$, but only the added set
contains $({\tt A}, j)$. The result is that ${\tt A}$ is present in
the set $s$.

In \cref{remove-noop-example}, instead of removing and re-adding {\tt
  A}, the first replica does nothing. Here, {\tt A} will \emph{not} be
present in the final state, as both the added and removed sets contain
only $({\tt A}, i)$.

We expect that undoing an operation brings us to the same state as if
it had never occurred, but this is not the case for OR-Sets. Doing and
undoing a $\n{remove}$ yields a different result from not removing at
all. After
removing an element from an OR-Set, there is in general no way to
revert to the previous state.

\subsection{Sets with removal: the PN-Set}
\label{pn-sets}

In a PN-Set, an element is present if it has been added more
times than it has been removed. The state is an unordered log of
operations ($\n{add}\;{\tt A}$ and $\n{remove}\;{\tt A}$), where an
element ${\tt A}$ is deemed present if there are more occurrences of
$\n{add}\;{\tt A}$ than $\n{remove}\;{\tt A}$.

On the same examples, the PN-Set gives a different result than the
OR-Set. In \cref{remove-undo-example}, in state $s$, the element {\tt
  A} has been added twice and removed twice, and is therefore
absent. Similarly, in state $s'$ of \cref{remove-noop-example}, the
element {\tt A} has been added once and removed once, and is therefore
absent. Unlike an OR-Set, all PN-Set operations are undoable: $\n{add}$ and
$\n{remove}$ perfectly cancel each other out.

However, the PN-Set allows unexpected extra states. Consider what happens when
executing the sequence $S₂$ from \cref{remove-undo-example}. After
performing $\n{add}\;{\tt A}$; $\n{remove}\;{\tt A}$;
$\n{remove}\;{\tt A}$, we reach a state in which ${\tt A}$ is present
-1 times: after performing $\n{add}\;{\tt A}$, the set will be empty.

This suggests an alternative representation of PN-Sets, as one copy of
the counter CRDT for each possible element, where $\n{add}$ and
$\n{remove}$ are implemented as increment and decrement.

\subsection{Sets with removal: the T-Set}

The extra states of a PN-Set arise because the counters can take
values other than 0 and 1. We can eliminate these states by using
modulo-2 counters instead of unbounded ones.

However, in a modulo-2 counter increment and decrement are the same
operation, so $\n{add}\;{\tt A}$ and $\n{remove}\;{\tt A}$ have the
same effect, toggling the membership of {\tt A}. We have eliminated
the extra states, but lost the distinction between $\n{add}$ and
$\n{remove}$.

\subsection{A trade-off}\label{tradeoff}

In choosing between these CRDTs, we face a trade-off: the OR-Set has
intuitive semantics for add and remove, but does not support general
undo. The PN-Set and T-Set do support undo, but work more like a tuple
of counters than a set, causing side-effects: extra states for PN-Sets
and nonstandard semantics for T-Sets.

More sophisticated CRDTs exhibit the same trade-off. For instance, the
Logoot-Undo CRDT for collaborative editing~\cite{logoot-undo} allows
all operations to be undone and redone, keeping count of how often
each operation has been undone. This supports general undo while
maintaining commutativity, but like PN-Sets it can be driven to a
state where an operation has been performed -1 times, and must be
redone to reach the empty state. The generic undo of Yu et
al.~\cite{generic-undo} also keeps undo counters, keeping track of
whether an operation has been undone an even or odd number of times
(like a T-Set).

The point of this note is that this trade-off is fundamental:
\emph{all undoable CRDTs are equivalent to a tuple of counters}.

\ifdefined\podcbrief

\else

\section{Formalising CRDTs}

To prove the theorem, we must first formalise undoable CRDTs. We
adopt a formulation of operation-based CRDTs close to Shapiro et al.'s
CmRDTs, except that we omit some details (e.g. message numbering) that
are not relevant here.

A CRDT consists of a set $S$ of \emph{abstract states} $s, t,
\dots$ with a distinguished \emph{initial state} $s₀$, and a
collection $P$ of \emph{primitive operations} $p, q, \dots$. We
assume that $P$ is finite, or equivalently that there is some upper
bound on the message length needed to communicate a single primitive
operation. $S$ may be infinite: there may be
infinitely many distinct states reachable by sequences of primitive
operations.

Each primitive operation $p ∈ P$ is a partial function from $S$ to
$S$.  That is, not all primitive operations need apply in all
states. To reduce parentheses, we write $s ⋅ p ⋅ q$ instead of
$q(p(s))$. We write $s ⋅ p ⋅ q\;\n{ok}$ when $s ⋅ p ⋅ q$ is
well-defined: that is, the operation $p$ applies in state $s$, and the
operation $q$ applies in state $s ⋅ p$. Note that $s ⋅ p ⋅ q\;\n{ok}$
implies $s ⋅ p\;\n{ok}$.

For simplicity, we assume that abstract states are neither impossible
nor redundant: we assume that distinct members of $S$ represent
logically distinct states, and all members of $S$ are reachable by
some sequence of primitive operations starting from $s₀$. If this isn't
true for a concrete implementation, we can choose the abstract states
$S$ by discarding unreachable states and picking one representative
among groups of logically equivalent states.

The property making states and primitive operations into a CRDT is
\emph{commutativity}: any two primitive operations that apply in the
same state commute. More formally, the structure is a CRDT if the
following axiom is satisfied (Definition 2.6 of Shapiro et al.~\cite{shapiro-crdt}):

\begin{axiom}[Commutativity]\label{ax-commute}
If $s ⋅ p\;\n{ok}$ and $s ⋅ q\;\n{ok}$, then $s ⋅ p ⋅ q\;\n{ok}$, $s ⋅
q ⋅ p\;\n{ok}$ and $s ⋅ p ⋅ q = s ⋅ q ⋅ p$.
\end{axiom}

Here, we're interested not in plain CRDTs but in \emph{undoable} ones,
which also satsify the following:

\begin{axiom}[Undoability]\label{ax-undo}
If $s ⋅ p\;\n{ok}$, then there exists some sequence of primitive
operations $q₁,\dots,q_n$ such that $s ⋅ p ⋅ q₁ ⋅ \dots ⋅ q_n$ is
well-defined and equals $s$.
\end{axiom}

Usually, a primitive operation $p$ will be undone using just
one operation $q$ (so $n = 1$), but we avoid assuming this.

\subsection{From operations to actions}

Rather than dealing with individual operations $p, q ∈ P$, it is more
convenient to consider the set $P^\ast$ of \emph{actions}. An action $a, b
∈ P^\ast$ is a finite sequence of primitive operations, which we apply to
states using the same notation: if $a = pq$, then $s⋅a = s⋅p⋅q$. We
write $ε$ for the empty action (so $s ⋅ ε = s$) and $ab$ for the
concatenation of $a$ and $b$ (so $s ⋅ ab = s ⋅ a ⋅ b$).

The axioms can be recast in terms of actions (see \cref{proofs}):
\begin{prop}[Commutativity of actions]\label{prop-commute}
If $s ⋅ a\;\n{ok}$ and $s ⋅ b\;\n{ok}$, then $s⋅ab$ and $s⋅ba$ are well-defined and equal.
\end{prop}
\begin{prop}[Undoability of actions]\label{prop-undo}
If $s ⋅ a\;\n{ok}$, then there exists some action $a^{-1}_s$ such that
$s ⋅ aa^{-1}_s = s$.
\end{prop}

\subsection{Equivalent CRDTs}\label{equivalent}

Next, we define what it means for two CRDTs to be equivalent. This is
more complicated that merely saying they have the same states and
primitive operations, because we want to view the counter CRDT (with
increment and decrement operations) as equivalent to a counter CRDT
that also exposes an ``increment twice'' operation.

So, we say that two CRDTs are equivalent if they have the same states
and both can implement each other's operations. Formally, a CRDT with
states $S₁$ and primitive operations $P₁$ is equivalent to one with
states $S₂$ and primitive operations $P₂$ if there is a one-to-one
(invertible) mapping $φ : S₁ → S₂$ as well as functions $ψ : P₁ → P₂^\ast$
and $ψ' : P₂ → P₁^\ast$ such that:
\begin{itemize}
\item $φ(s₀) = s₀'$
\item If $s ⋅ p\;\n{ok}$, then $φ(s)⋅ψ(p) = φ(s⋅p)$
\item If $s' ⋅ p'\;\n{ok}$, then $φ^{-1}(s') ⋅ ψ'(p) = φ^{-1}(s'⋅p')$
\end{itemize}

In other words, two equivalent CRDTs are two representations for the
same data structure, and we can apply operations to states in either
representation. Two machines using equivalent CRDTs can coexist on the
same network: as long as they translate their messages back and forth
using $ψ$ and $ψ'$ neither will be able to tell that the other is
using a different internal representation. For instance, the two
representations of PN-Sets in \cref{pn-sets} (as unordered logs and as
per-element counters) are equivalent.

\subsection{The tuple construction}

Given two CRDTs $A$ and $B$, we can combine both into a single CRDT
using a straightforward construction. The states of the combined CRDT
are pairs $(s_A, s_B)$ of a state of $A$ and a state of $B$, and all
the primitive operations of $A$ and $B$ are primitive operations of
the combined CRDT, with the operations of $P_A$ acting on the left
half of the state and the operations of $P_B$ acting on the right.

Effectively, the combined CRDT acts as two independent CRDTs, one
implementing $A$ and one implementing $B$. This construction is not
limited to just two CRDTs: we may form tuples of $n$ CRDTs in the same
way.

\bigskip\noindent
This gives us enough ingredients to formally state the theorem:
\begin{theorem*}
Every undoable CRDT is equivalent to a tuple of counter and modulo counter CRDTs.
\end{theorem*}

\subsection{The group of actions}

The proof of this theorem relies on some classical group theory,
applied to the \emph{group of actions} of an undoable CRDT.

First, given any undoable CRDT and a state $s ∈ S$, we define the
relation $≡_s$ on actions so that $a ≡_s b$ whenever $s ⋅ a\;\n{ok}$,
$s ⋅ b\;\n{ok}$ and $s ⋅ a = s ⋅ b$. This is a \emph{partial
  equivalence relation}: it is transitive and symmetric, but not
reflexive since $a ≡_s a$ is not true in general, but only when $s ⋅
a\;\n{ok}$.

Now, given $a₁ ≡_s b₁$ and $a₂ ≡_s b₂$, Commutativity tells us that
all of $a₁, a₂, b₁, b₂$ commute with each other (since all apply in
state $s$), and so:
$
s⋅a₁a₂ = s⋅b₁a₂
= s⋅a₂b₁
= s⋅b₂b₁
= s⋅b₁b₂
$. Therefore:
\begin{fact}\label{concat-binary}
If $a₁ ≡_s b₁$ and $a₂ ≡_s b₂$, then $a₁a₂ ≡_s b₁b₂$
\end{fact}
By applying Commutativity with $a = b$, we learn that any valid action
can be done twice (since it may be performed independently by two
replicas, which later merge):
\begin{fact}
If $s⋅a\;\n{ok}$, then $s⋅aa\;\n{ok}$
\end{fact}
Combining this with Undoability, we learn that actions can be undone
twice:
\begin{fact}
If $s⋅a\;\n{ok}$, then $(s⋅a)⋅a^{-1}_sa^{-1}_s\;\n{ok}$
\end{fact}
But since $s⋅aa^{-1}_s = s$, this has the surprising consequence that
actions can be undone before they are performed:
\begin{fact}
If $s⋅a\;\n{ok}$, then $s⋅a^{-1}\;\n{ok}$
\end{fact}
Commutativity then tells us that $a$ and $a^{-1}_s$ commute:
\begin{fact}\label{inverses}
If $s⋅a\;\n{ok}$, then $a^{-1}_sa ≡_s aa^{-1}_s ≡_s ε$.
\end{fact}

These facts mean that we can form a group $G_s = P^\ast/{≡}_s$ of the
equivalence classes of $≡_s$: concatenation is a binary operation on
$P^\ast/{≡}_s$ thanks to \cref{concat-binary}, and inverses exist thanks to
\cref{inverses}. In other words, members of the group $G_s$ are
denoted by actions that apply to state $s$, with two actions denoting
the same member of the group if they yield the same result when
applied to $s$. Since all members of this group commute, the group is
\emph{abelian}.

\subsection{$G_s$ is finitely generated}

Just as actions are built out of a finite set $P$ of primitive
operations, elements of $G_s$ are built out of a finite set $P/{≡}_s$
of \emph{generators}. To prove this, we first note that any action
that can be performed later can be performed now. If $s⋅ab\;\n{ok}$,
then by \cref{inverses} $s⋅a^{-1}_s\;\n{ok}$, and so
$s⋅a^{-1}_sab\;\n{ok}$ by Commutativity, whence:
\begin{fact}\label{ab-bok}
If $s⋅ab\;\n{ok}$, then $s⋅b\;\n{ok}$.
\end{fact}
Therefore, given any action $a = p₁p₂\dots p_n$ such that
$s⋅a$, we have that $s⋅p_i\;\n{ok}$ for all $1 ≤ i ≤ n$: first
by noting $s⋅p₁p₂\dots p_i\;\n{ok}$, and then by applying
\cref{ab-bok}. So, each element of $G_s$ can be written as the
concatenation of a sequence of primitive operations $p$ that apply in
state $s$: in other words, $G_s$ is \emph{generated} by
$P/{≡}_s$.

So, in any undoable CRDT, the actions available from any state have
the structure of a \emph{finitely generated abelian group}.

\subsection{An old theorem}

To show that undoable CRDTs are equivalent to tuples of counters,
it's enough that they have isomorphic groups of actions,
thanks to the following (proof in appendix):
\begin{prop}\label{prop-iso}
If the groups of actions $G_{s₀}$ and $G_{s₀'}$ of two CRDTs are
isomorphic, then the CRDTs are equivalent.
\end{prop}

The groups of actions of any counter CRDTs is a \emph{cyclic group}:
either $ℤ$, the group of integers with addition (for unbounded
counters), or $ℤ_n$, the group of integers with addition modulo $n$
(for counters modulo $n$). The group of actions of a tuple of $n$
CRDTs is given by an action for each of the $n$ components of the
tuple, composed pointwise: this is the \emph{direct sum} of their
groups of actions.

Since the group of actions of an undoable CRDT is a finitely generated
abelian group, our theorem follows from an old result, the
\emph{fundamental theorem of finitely generated abelian groups}:

\begin{theorem*}[Poincaré 1900; Kronecker 1870, Noether 1926]
Every finitely generated abelian group is isomorphic to the direct sum
of finitely many cyclic groups.
\end{theorem*}

See e.g. Rotman~\cite[p.318]{rotman} for a proof, or
Stillwell~\cite[p.175]{stillwell} for a proof and some history.

\fi

\section{Discussion}

This characterisation of undoable CRDTs has a number of immediate
consequences, including:
\begin{description}
\item[All operations are always valid] For
  instance, an undoable CRDT cannot represent a nonnegative counter,
  in which decrement is available only in nonzero states.
\item[Negative states always exist] For any action $a$, there is some
  state $s$ in which applying $a$ will bring us back to the initial
  state $s_0$.
\end{description}

In the specific example of set CRDTs, we see that the trade-off
described in \cref{tradeoff} is unavoidable: in any CRDT with
$2^n$ states representing presence or absence of $n$ elements, one of
the following must be true:
\begin{itemize}
\item Some operations are not undoable (like OR-Set)
\item There are an infinite number of extra states, beyond the $2^n$
  states representing membership (like PN-Set)
\item All operations must be cyclic, undoing themselves after some
  number of iterations (like T-Set)
\end{itemize}

In light of this, designers of distributed data structures must 
limit themselves to tuples of counters, accept that some operations
will not be fully undoable, or use something other than CRDTs.

\begin{acks}
Thanks to KC Sivaramakrishnan for the conversation that originally led
to this paper, to Martin Kleppmann for detailed feedback, and to
Louise Doran and Anil Madhavapeddy for useful comments.
\end{acks}

\bibliographystyle{abbrvnat}
{
\raggedright
\bibliography{references}
}

\ifdefined\podcbrief
\else

\appendix
\newpage
\section{Additional proofs}\label{proofs}

\begin{proof}[Proof of \cref{prop-commute}] First, we show that if $s ⋅
p\;\n{ok}$ and $s ⋅ a\;\n{ok}$ then $pa ≡_s ap$, by induction on the
length of $a$. If $a = ε$, then the result follows. Otherwise $a =
qb$. By \cref{ax-commute}, $s⋅p⋅q$ and $s⋅q⋅p$ are defined and
equal. Thus, we have both $(s⋅q)⋅p\;\n{ok}$ and $(s⋅q)⋅b\;\n{ok}$, so
the inductive hypothesis gives $pb ≡_{s⋅q} bp$ or equivalently $qpb
≡_s qbp$. Gluing these together, we get:
\[
pa = pqb ≡_s qpb ≡_s qbp = ap
\]
Next, we use this fact to prove that if $s⋅a\;\n{ok}$ and
$s⋅b\;\n{ok}$ then $ab ≡_s ba$ by similar induction on the length of
$b$. If $b = ε$ then the result is again trivial. Otherwise, $b = pc$
and by above, $pa ≡_s ap$. Thus, we have both $(s⋅p)⋅a\;\n{ok}$ and
$(s⋅p)⋅c\;\n{ok}$, so the inductive hypothesis gives $ac ≡_{s⋅p} ca$
or equivalently $pac ≡_s pca$, leading to:
\[ab = apc ≡_s pac ≡_s pca = ba \qedhere\]
\end{proof}

\begin{proof}[Proof of \cref{prop-undo}]
Again, we proceed by induction on $a$. If $a=ε$, then $a^{-1}_s = ε$
suffices. Otherwise $a = bp$, so we choose $a^{-1}_s = q₁q₂\dots
q_nb^{-1}_s$, where $q_i$ are those given by \cref{ax-undo} for state
$s⋅b$. Then:
\[s⋅bpq₁q₂\dots q_nb^{-1}_s = s⋅bb^{-1}_s = s \qedhere\]
\end{proof}

\begin{proof}[Proof of \cref{prop-iso}]
Given a CRDT with states $S₁$ and operations $P₁$ and one with states
$S₂$ and operations $P₂$, suppose that an isomorphism $ψ$ exists
between $G_{s₀}$ and $G_{s'₀}$. We define the mappings $φ : S₁ → S₂,
φ^{-1} : S₂ → S₁$ as follows:
\begin{align*}
φ(s) &= s'₀ ⋅ ψ(a) &&\text{ for some $a$ such that $s₀ ⋅ a = s$} \\
φ^{-1}(s') &= s₀ ⋅ ψ^{-1}(a') &&\text{ for some $a'$ such that $s₀' ⋅ a' = s'$}
\end{align*}
Such actions $a, a'$ must exist because all states are reachable in
both CRDTs. If several are possible, the choice of $a, a'$ does not
matter, since $ψ$ respects $≡_{s₀}$ and so must map all such $a$ to
equivalent actions.

These functions are inverses:
\begin{align*}
φ^{-1}(φ(s)) &= φ^{-1}(s'₀ ⋅ ψ(a)) = s₀ ⋅ ψ^{-1}(a') \\
\text{where} & \quad s₀' ⋅ a' = s₀' ⋅ ψ(a) \\
 & \quad s₀ ⋅ a = s
\end{align*}
Since $a' ≡_{s₀'} ψ(a)$, \[ψ^{-1}(a') ≡_{s₀} ψ^{-1}(ψ(a)) ≡_{s₀} a\]
So, $s₀ ⋅ ψ^{-1}(a') = s₀ ⋅ a = s$. The proof that $φ(φ^{-1}(s)) = s$
is identical.

From $ψ$, we get a mapping $P₁ → P₂^\ast$ (and likewise $ψ^{-1}$ gives a
mapping $P₂ → P₁^\ast$). To prove the CRDTs equivalent, we must show that
these satisfy the three conditions from \cref{equivalent}:
\begin{itemize}
\item $φ(s₀) = s₀' ⋅ ψ(a)$ where $s₀ ⋅ a = s₀$. But since $a ≡_{s₀}
  ε$, $ψ(a) ≡_{s₀} ε$ and so $φ(s₀) = s₀'$.
\item Suppose $s ⋅ p\;\n{ok}$. Then, for some $a$ where $s₀ ⋅ a = s$,
\[
φ(s) ⋅ ψ(p) = s₀ ⋅ ψ(a) ⋅ ψ(p) = s₀ ⋅ ψ(ap) = φ(s⋅p)
\]
\item As above \qedhere
\end{itemize}
\end{proof}
\fi

\end{document}